\documentclass[pdftex,twocolumn,amsmath,amssymb,epjc3]{svjour3}          

\RequirePackage[T1]{fontenc}

\smartqed  

\RequirePackage{graphicx}
\RequirePackage{mathptmx}      
\RequirePackage{flushend}
\RequirePackage[numbers,sort&compress]{natbib}
\RequirePackage[colorlinks,citecolor=blue,urlcolor=blue,linkcolor=blue]{hyperref}

\usepackage{mathtools}

\journalname{ }

\begin{document}

\title{Lattice model for self-folding at the microscale}

\author{T. S. A. N. Sim\~oes\thanksref{e1,addr1,addr2}
        \and
        H. P. M. Melo\thanksref{e2,addr1} 
		  \and
		  N. A. M. Ara\'ujo\thanksref{e3,addr1,addr3} 
}

\thankstext{e1}{e-mail: pg38008@alunos.uminho.pt}
\thankstext{e2}{e-mail: hpmelo@fc.ul.pt}
\thankstext{e3}{e-mail: nmaraujo@fc.ul.pt}

\institute{Centro de F\'{i}sica Te\'{o}rica e Computacional, Faculdade de Ci\^encias, Universidade de Lisboa, 1749-016 Lisboa, Portugal\label{addr1}
          \and
			 Centro de F\'{i}sica das Universidades do Minho e do Porto, Campus de Gualtar, 4710-057 Braga, Portugal\label{addr2}
          \and
          Departamento de F\'isica, Faculdade de Ci\^encias, Universidade de Lisboa, 1749-016 Lisboa, Portugal\label{addr3}
}

\date{Received: date / Accepted: date}

\abstractdc{Three-dimensional shell-like structures can be obtained spontaneously at the microscale from the self-folding of 2D templates of rigid panels. At least for simple structures, the motion of each panel is consistent with a Brownian process and folding occurs through a sequence of binding events, where pairs of panels meet at a specific closing angle. Here, we propose a lattice model to describe the dynamics of self-folding. As an example, we study the folding of a pyramid of $N$ lateral faces. We combine analytical and numerical Monte Carlo simulations to find how the folding time depends on the number of faces, closing angle, and initial configuration. Implications for the study of more complex structures are discussed. }

\maketitle

\section{Introduction}

Folding is considered a promising strategy to design shape-changing materials with fine tuned mechanical, electrical, and optical properties~\cite{shyu2015kirigami,blees2015graphene,liu2018nano,zhang2015mechanically}. The idea is to act over a single degree of freedom to promote a reversible change in shape, as in the case of the Miura-Ori fold~\cite{miura1985method,mahadevan2005self,schenk2013geometry}. The folding trajectories are usually deterministic and the final folded configuration unique. However, with technology putting pressure to go to smaller and smaller scales as, for example, in encapsulation and soft robotics~\cite{fernandes2012self,shim2012buckling,filippousi2013polyhedral,felton2014method}, new challenges are posed. At the microscale, thermal fluctuations are no longer negligible and so the folding process is stochastic. Experimental and numerical results show that the final folded configuration is no longer unique, but rather dependent on the experimental conditions~\cite{pandey2011algorithmic,araujo2018finding,dodd2018universal,azam2009compactness}. 

Thermal fluctuations have also advantages. They can drive the folding process and, if the templates are properly designed, there is no need to control any degree of freedom. Recently, the spontaneous folding of a pyramid from a planar template consisting of rigid micron-size panels connected through flexible hinges has been studied using molecular dynamics simulations~\cite{Melo2020}. The numerical results suggest that, when thermal fluctuations dominate, the motion of individual panels is well described by a Brownian process and folding evolves through a sequence of binding events between pairs of faces, which correspond to first-passage processes.

The first binding event occurs when the first pair of lateral faces meet at the closing angle $\phi$. 
The characteristic time is the first binding time $T_{F}$ and can be estimated by mapping the problem into a two-dimensional first-passage process \cite{Melo2020}. The remaining $N-2$ faces move freely until they also have the same closing angle. Therefore, the subsequent binding events can all be mapped to a one-dimensional first-passage processes \cite{Melo2020}. When the last face binds, the folding process is complete. The time between the first and last binding is called the last binding time $T_{L}$. The total folding time $T$ is then $T=T_{F}+T_{L}$.

Due to the stochastic nature of the folding process, the timescale of binding is typically much larger than the one of thermal jiggling of the individual panels. 
Thus, to observe the complete folding, one needs very long molecular dynamics simulations. In order to access the folding timescales efficiently, here, we propose
a lattice model. We also consider the folding of a pyramid with $N$ lateral faces. 
We show that the dynamics of folding can be mapped into a set of random walks on a lattice, each one representing
 a lateral face, and the binding events occur when two random walks visit a predefined set of lattice sites simultaneously. We show that our model 
is consistent with existing results and reproduces the relation between 
folding time and number of faces $N$, reported previously~\cite{Melo2020}. The proposed lattice model  
allows us to explore the folding dynamics for different structures and initial conditions.
The model is generic and can be generalized to simulate the folding of more complex structures, by including new interaction 
zones and/or changing the interaction rules.

The paper is organized as follows. In sect.~\ref{sec_model}, we introduce the lattice model and we define the first and last 
binding times. In sect.~\ref{sec_results} we present numerical and analytical results
for the total folding time. We draw some conclusions in sect.~\ref{sec_conclusions}.

\begin{figure}[t]
\includegraphics[width=8.5cm]{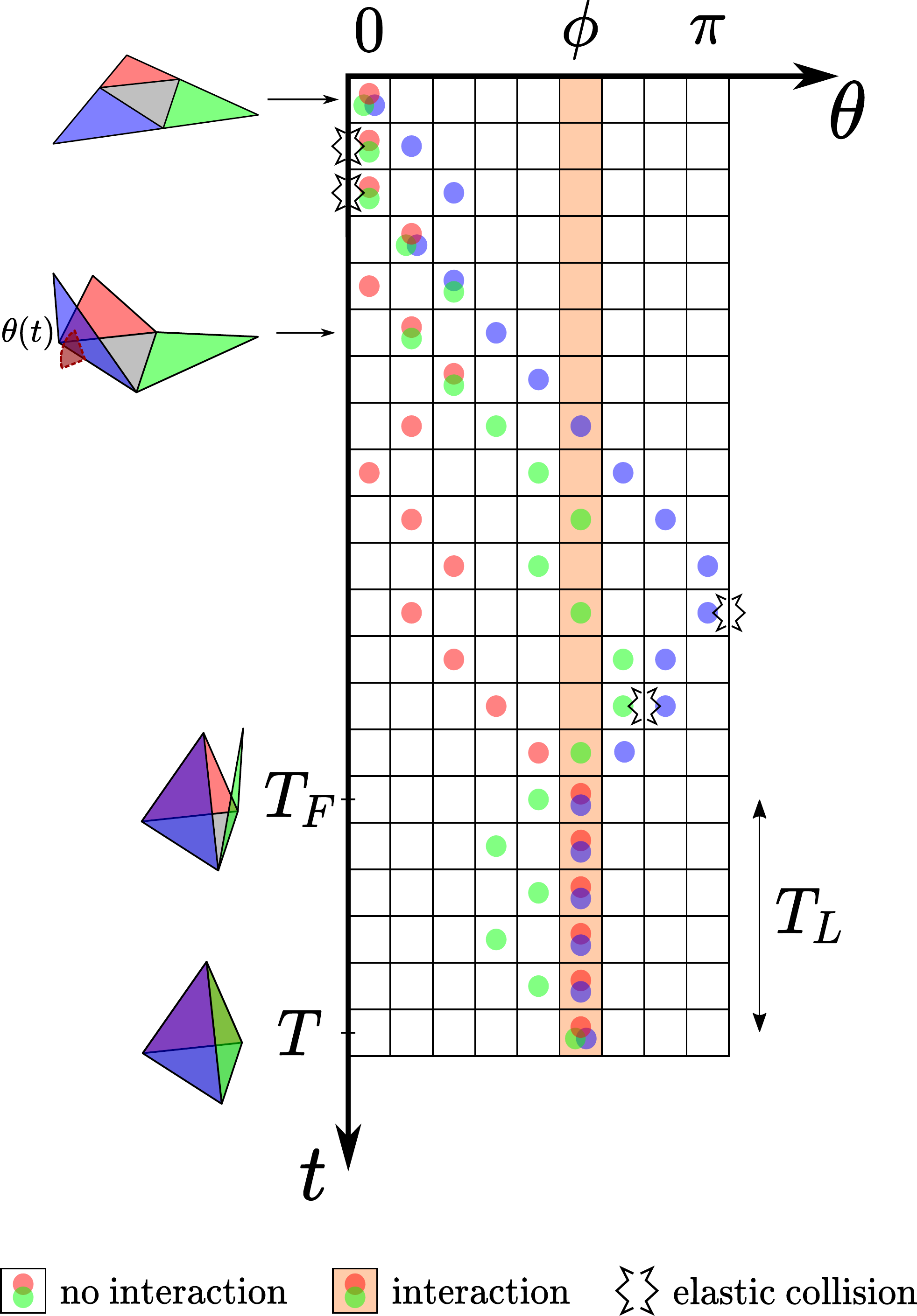}
\caption{ Schematic representation of the folding of a pyramid of three lateral faces and closing angle $\phi$, starting from a flat template, and the corresponding dynamics on the lattice model. The horizontal axis represents the angle $\theta$ of each face at a certain time $t$, in the domain $\theta\in\left[0,\pi\right]$ with reflective boundaries, and the vertical axis is the time $t$. The motion of each face is mapped into the movement of a random walker in a 1D. The first binding time $T_{F}$ is the time at which the first pair of walkers meet at the closing angle $\phi$ and bind together. The folding time $T$ is when all walkers meet at the closing angle, thus completing the folding process. The last binding time $T_{L}$ is defined to be the time interval between first binding and the folding time, so $T_{L}=T-T_{F}$. The excluded volume interaction between walkers and with boundaries is described by elastic collisions as depicted. The corresponding configuration for four times ($t=0$, $0<t<T_{F}$, $t=T_{F}$ and $t=T$) is also shown.}
\label{fig_1}
\end{figure}

\section{Model}
\label{sec_model}

We map the spontaneous folding of a regular pyramid with $N$ lateral faces into a set of random walks on a one-dimensional lattice. 
Each face $i$ is a random walk and the angle $\theta_{i}$ relative to the base is given by the position on a 1D lattice of $l$ sites.  
To study the folding of a regular pyramid without misfolding, we consider that the domain of the lattice ranges from $\theta\in\left[0,\pi\right]$, such that the site $j$ corresponds to an angle in the range $\left[j\frac{\pi}{l},(j+1)\frac{\pi}{l}\right]$, see model scheme in Fig.~\ref{fig_1}. 

We consider a rejection-free kinetic Monte Carlo scheme \cite{andersen2019practical,stamatakis2014kinetic,sickafus2007radiation}, where, at each time step, a randomly selected walker tries to hop to one of its neighbors. Time is incremented by $\Delta t = \frac{\Delta\theta^{2}}{2D}\frac{1}{N_{F}}$, where $\Delta\theta=\frac{\pi}{l}$ is the lattice spacing, $D$ is the angular diffusion coefficient of the lateral faces, and $N_{F}$ is the number of walkers that are still free.
For simplicity, we consider an unbiased folding, and thus the hopping probability is the same in both directions. 

The binding between any two lateral faces occurs when two random walkers first meet at a specific site corresponding to the closing angle $\phi$. 
Additionally, when two faces are in the region beyond the closing angle ($\theta>\phi$) there is a geometrical constraint that they can not overpass each other. 
To include this interaction, we assume that in the region $\theta\in\left]\phi,\pi\right]$, random walkers interact through excluded volume: If a walker tries to hop into an already occupied site, it goes back to the original site.

\section{Results}
\label{sec_results}

For each number of random walkers $N$, we performed $10^4$ independent samples, starting with all walkers at the origin of the 
lattice, which correspond to the open template shown in Fig.~\ref{fig_1}.
Figure~\ref{fig_2} shows the average folding time $\langle T \rangle$, where we recover the non-monotonic dependence on the number of lateral faces $N$, reported previously~\cite{Melo2020}. 
This behavior stems from a different functional dependence on $N$ of the mean first binding time $\langle T_{F} \rangle$ and the
 mean last binding time $\langle T_{L} \rangle$, as discussed below.

\subsection{First binding time}
 
\begin{figure}[t]
\includegraphics[width=8.5cm]{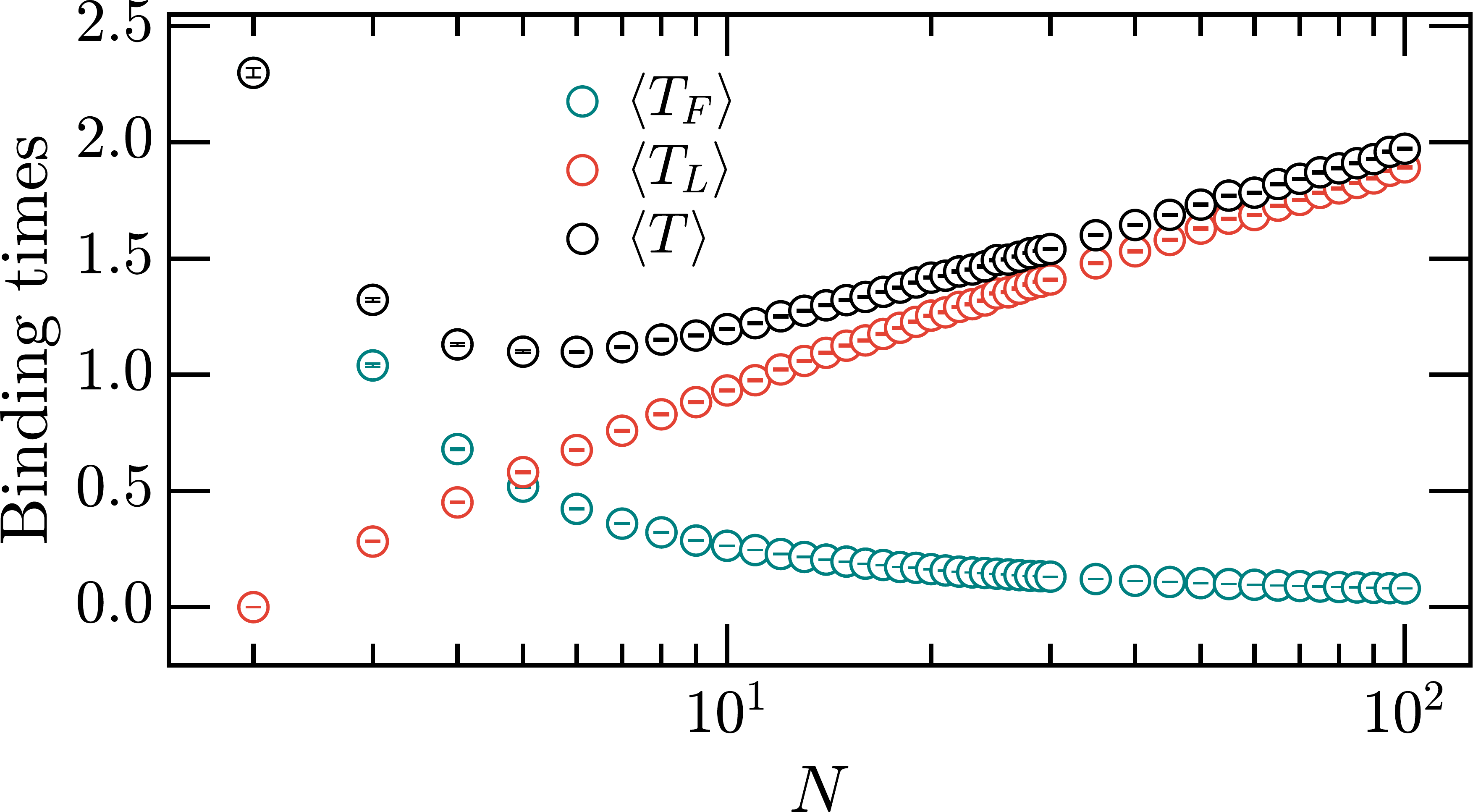}
\caption{\textbf{Dependence of the folding time on the number of lateral faces $N$.} $\langle T_{F} \rangle$, $\langle T_{L} \rangle$ and $\langle T \rangle$ are the average first binding, last binding and folding time, respectively. Time is in units of Brownian time, i.e. $\frac{\pi^2}{2D}$, where $D$ is the diffusion coefficient of each face. 
We started from a planar condition, where $\theta_i=0$ for all faces $i$, and a fixed closing angle to $\phi=\frac{2}{3}\pi$.  
Results are averages over $10^{4}$ independent samples on a lattice of $l=181$ sites and the error bars are given by the standard error.}
\label{fig_2}
\end{figure}

\begin{figure}[t]
\includegraphics[width=8.5cm]{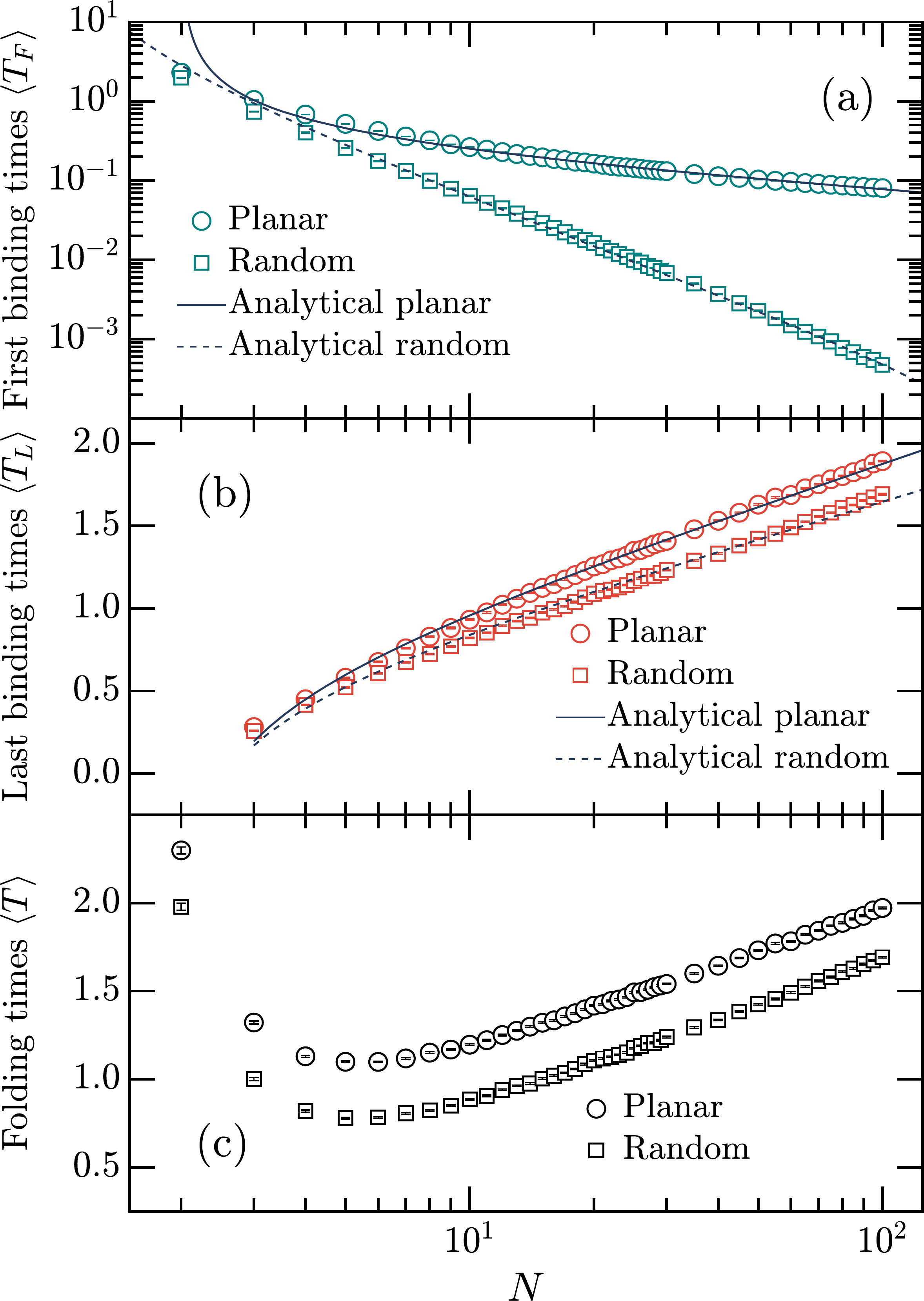}
\caption{\textbf{Dependence on the initial conditions.} 
We consider two initial conditions: all faces start at the origin (planar) 
or uniformly distributed (random).
(a) First binding time as a function of $N$ for the two initial conditions. The solid and dashed lines are given by $T_F=\omega_{P_0} + \omega_P/\ln(N(N-1)/2)$ where $\omega_{P_0}=-0.0632\pm 0.0003$ and $\omega_P=1.199\pm 0.002$ for the planar, and $T_F=2\omega_R/N(N-1)$ with $\omega_R=2.830\pm 0.007$ for the random, all obtained by least mean square fitting. 
(b) Last binding time as a function of $N$. Asymptotically for both initial conditions time scales logarithmically with $N$, as predicted by Eq.~\ref{eq.last_meeting_2}. 
The solid and dashed lines given by $T_L=\tau_{P_0}+\tau_P\ln(N-2)$ where $\tau_{P_0}=0.196\pm 0.002$ and $\tau_{P}=0.3662\pm 0.0006$ for the planar,
and $T_L=\tau_{R_0}+\tau_R\ln(N-2)$ where $\tau_{R_0}=0.171\pm 0.002$ and $\tau_{R}=0.3217\pm 0.0006$ for the random, all obtained by least square fitting.
(c) Total folding time. All times are in units of Brownian time and a fixed closing angle of $\phi=\frac{2}{3}\pi$. Results are averages over $10^{4}$ independent samples on a lattice of $l=181$ sites and the error bars are given by the standard error.}
\label{fig_3}
\end{figure}

We first estimate the dependence of $\left \langle T_{F} \right \rangle$ on $N$.
For a pyramid with $N$ lateral faces, the time that the first two faces bind can be estimated in the following way. Let us define $g(t)$ as the probability density function that a pair of Brownian particles meet at an angle $\phi$ at time $t$ for the first time. 
Assuming there is no correlation between the movement of the \mbox{$N_{p}=\binom{N}{2}=\frac{N(N-1)}{2}$} pairs, for each sample, $T_F=min(t_1,t_2,t_3...t_{N_{p}})$, where $t_i$ is a random number drawn from the distribution $g(t)$, for each pair $i$. 
From the theory of order statistics~\cite{arnold1992first} the expected time for the first binding is,
\begin{equation}
\left \langle T_F(N_{p}) \right \rangle = N_{p}\int_0^\infty{t g(t) G(t)^{N_{p}-1} dt}, \label{eq.first_meeting_0}
\end{equation}
where $G(t)=\int_t^\infty g(t')dt'$ is the survival function. Using partial integration we obtain,
\begin{equation}
\left \langle T_F(N_{p}) \right \rangle = \int_0^\infty{e^{N_{p}\ln G(t)} dt}. \label{eq.first_meeting}
\end{equation}
Equation~\ref{eq.first_meeting} shows that, since $\ln G(t)\le 1$, the first binding time $T_F$ is a monotonic decreasing function of $N$,
as also observed with the lattice model (see Fig.~\ref{fig_2}). The value of $\left \langle T_F \right \rangle$ for $N\to \infty$ 
can be estimated for different geometries and initial conditions
~\cite{yuste1996order,weiss1983order,holcman2014narrow,singer2006narrow,yuste1997escape}. 
For all faces starting at $\theta_i(0)=0$ (planar initial condition), it was shown that $\left \langle T_{F} \right \rangle \sim 1/\ln(N_{p})$~\cite{weiss1983order,basnayake2019asymptotic},
which is in agreement with the simulation results shown in Fig.~\ref{fig_3}(a).

For the random initial condition, all walkers start at a lattice site selected uniformly at random. 
Since there is a chance of two faces to start close to $\phi$, we can assume that $G(t)$ decays rapidly with $t$,
and for large values of $N$ the most significant contribution for the integral in Eq.~\ref{eq.first_meeting} is from $t$ near zero.
Therefore, using the approximation $\ln G(t)\approx \ln(G_0) + \frac{G'_0}{G_0}t$, where $G_0=G(t=0)=1$ and $G'_0=\frac{dG}{dt}(t=0)$, 
the first binding time can be calculated as 
\begin{equation}
\left \langle T_F(N) \right \rangle \approx -\frac{2}{N(N-1)G'_0}. \label{eq.first_meeting_random}
\end{equation}

Since $G'_0<0$, Eq.~\ref{eq.first_meeting_random} implies that the first binding time decays with $1/N(N-1)$ in perfect agreement 
with the simulation results shown in Fig.~\ref{fig_3}(a).

\subsection{Last binding time}

After the first closing, the two faces that form a bond stay fixed at the angle $\phi$, while the remaining $N-2$ faces move 
until they reach the same angle for the first time. To determine  $\left \langle T_{L} \right \rangle$, 
we calculate the average time for the binding of the last face.
From a first-passage time probability $f(t)$, we have $T_L=max(t_1,t_2,t_3...t_{N-2})$, where the average time is given by
\begin{equation}
\left \langle T_L(N) \right \rangle = (N-2)\int_0^\infty{t f(t) \left[1-F(t)\right]^{(N-2)-1} dt}, \label{eq.last_meeting}
\end{equation}
where $F(t)=\int_t^\infty f(t')dt'$. We assume an uniform distribution of $\theta_i<\phi$ for all the $N-2$ free faces at the moment of the first binding.
For large values of $t$, the distribution of first-passage times is well described
by $f(t)\approx e^{-t/\tau_L}$, with $\tau_L=4 \phi^2/D\pi^2$~\cite{redner2001guide}, and the average time for the last binding is
\begin{equation}
\left \langle T_L(N) \right \rangle = \tau_L \sum_{i=1}^{N-2} \frac{1}{i}.\label{eq.last_meeting_2}
\end{equation}
For large $N$, $\left \langle T_L(N) \right \rangle \approx \tau_L \ln(N-2) + \gamma\tau_L $, where $\gamma$ is the Euler-Mascheroni 
constant. In Fig.~\ref{fig_3}(b) we show that the last binding time grows with the logarithm of $N$ as predicted 
by our analytical calculations for both initial conditions.

For low values of $N$, the first binding time is large and, therefore, after the first binding 
the $N-2$ free faces are uniformly distributed, $T_L$ is the same for the planar and random initial condition (see Fig.~\ref{fig_3}(b)). 
However, for large values of $N$, we have a fast first binding and the remaining $N-2$ free faces will still be close to $\theta_i=0$ for the planar initial condition.
This is consistent with the fact that $T_L$ is larger for the planar than the random initial condition (see Fig.~\ref{fig_3}(b)).

Since the average folding time is the sum between $\langle T_{F} \rangle$ and $\langle T_{L} \rangle$, the different functional dependencies for these two quantities explains the non-monotonic behavior of $\langle T \rangle$ shown in Fig.~\ref{fig_3}(c) for both initial conditions, in line with previous results obtained using molecular dynamic simulations~\cite{Melo2020}.

\subsection{Closing angle}

\begin{figure}[t]
\includegraphics[width=8.5cm]{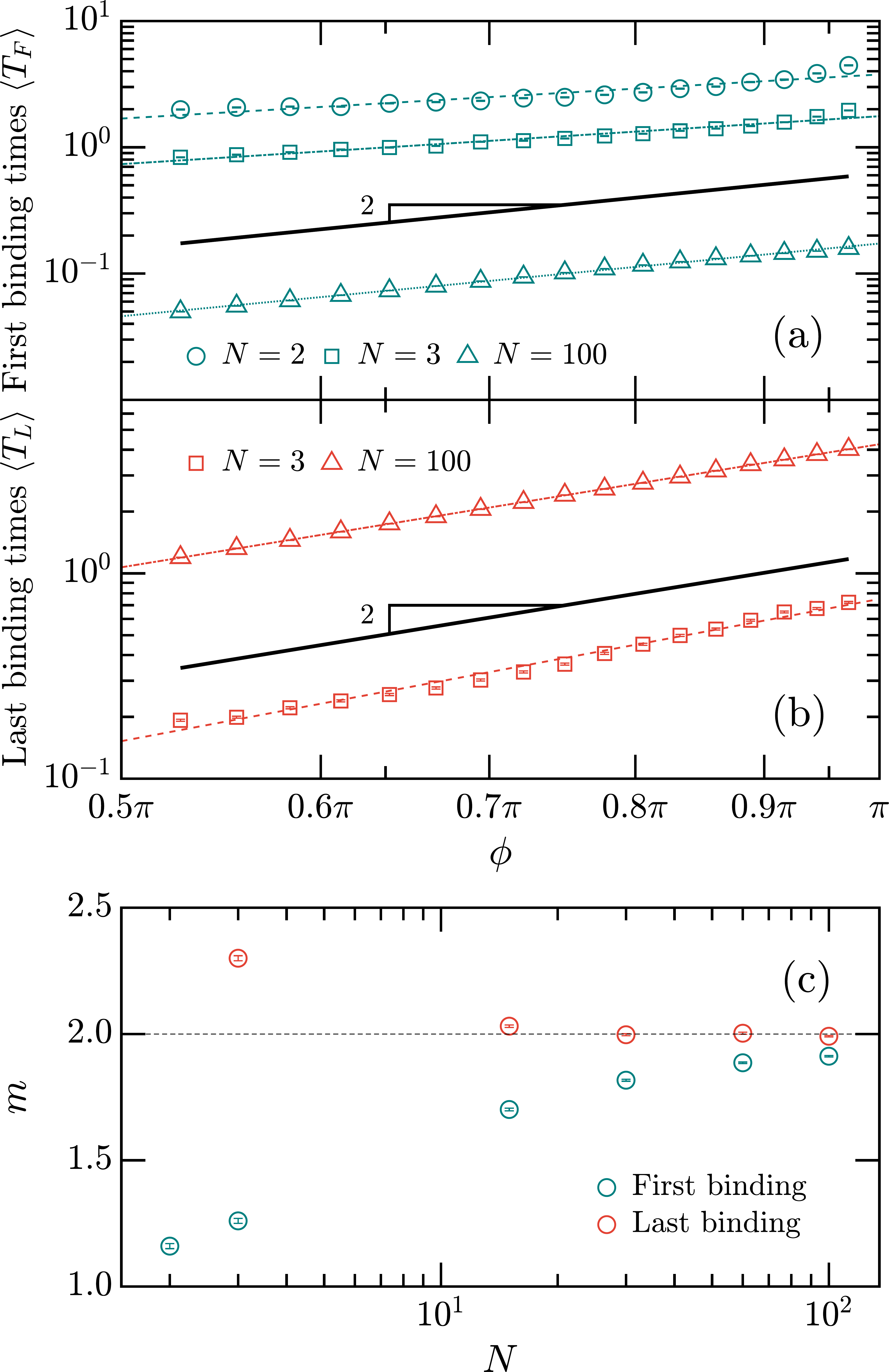}
\caption{\textbf{Time dependence with the closing angle $\phi$.} 
(a) First binding time $\langle T_{F} \rangle$ as a function of $\phi$ for $N=\{2,3,100\}$.
(b) Last binding time $\langle T_{L} \rangle$ as a function of $\phi$ for $N=\{3,100\}$.
In (a) and (b) the data was fitted by a power law  $\langle T_{L/F} \rangle\sim\phi^{m}$, represented by the dashed lines. 
From the theory of first-passage processes, the time scale should be proportional to the square of the domain size,
however since the adsorbing site is not on the boundary of the lattice we can have deviations from that scaling, corresponding to 
$m\neq 2.0$. We show in (c) that for both the first and last binding times, $m$ converges to $2.0$ as we increase $N$, since for large 
$N$ the first face to reach $\phi$ closely follows a straight path~\cite{lawley2020probabilistic}. All times are in units of Brownian time and $\phi\in\left[\frac{19\pi}{36},\frac{35\pi}{36}\right]$. Results are averages over $10^{4}$ independent samples with planar initial conditions on a lattice of $l=181$ sites and the error bars are given by the standard error.}
\label{fig_4}
\end{figure}

Since the overall dynamics of all faces is diffusive, we would expect that the time scale that controls the binding process is the Brownian time $\phi^2/2D$. Figure \ref{fig_4} shows simulation results for the average first (a) and last (b) binding times as a function of $\phi$, for different values of the number of faces $N$. The solid and dashed lines are given by fitting the simulation data to a power law $\langle T_F \rangle=A_F\phi^{m_F}$ and $\langle T_L \rangle=A_L\phi^{m_L}$, using least mean square method, with fitting parameters $(A_F,m_F)$ and $(A_L,m_L)$.
However, the results show that the binding time tends to be proportional to $\phi^{2}$ only for large values of $N$. In particular, for $N=2$ and $N=3$, we see that there is 
a slight deviation from the quadratic scaling. This deviation is due to the fact that 
the first binding is a two-dimensional first-passage process where 
the binding site is located inside the domain, and the time scale is controlled by the size of the boundary. However, for large values of $N$ the path for the first binding will be close to a straight line~\cite{lawley2020probabilistic}, recovering the timescale $\phi^2/2D$ as shown 
in Fig.~\ref{fig_4}(c).

\section{Discussion and conclusions}
\label{sec_conclusions}

We proposed a lattice model to simulate the spontaneous folding of a pyramid
at the microscale, driven by thermal fluctuations. We map the angular motion of each face into a random walk on a lattice. The 
pairwise binding of faces corresponds to having two walkers at a specific lattice site. 
We recover a recent result obtained from Molecular Dynamics simulations, namely, that the average folding time is a non-monotonic
function of the number of lateral faces $N$. The folding dynamics involves two types of first-passage processes: 
The first binding, where the first two faces bind together; and the last binding, where the last 
face binds. We show that the first binding corresponds to a two-dimensional first-passage process,
 where two random walkers have to bind for the first time at a specific
lattice site. After that, the remaining free walkers bind through a sequence of one-dimensional first-passage processes. 
The total folding time is the sum of the first and last binding times.

Describing folding as a sequence of binding events, we demonstrated that 
the characteristic time of the first binding has to decrease with $N$, while for the 
last binding it increases. It is the balance between these two processes that leads to a non-monotonic dependence 
on the number of faces. We show that this non-monotonic dependence is robust to changes on the initial conditions 
or value of the closing angle. Although the initial condition could affect both first and last binding times, it only affects significantly the 
first binding time and only for large $N$. For a small number of lateral faces the binding times slightly deviates from a quadratic dependence on $\phi$, however for large $N$ the quadratic dependence is recovered, as expected for a diffusive process.

The lattice model allows to evaluate not only the folding time, but also the yield of more 
complex structures by extending the domain and including new closing angles. 
For example, an additional closing angle at $-\phi$ corresponds to the possibility of folding in two different sides of the base of the pyramid, with possible misfolding for a pyramid with $N>3$, 
where the final folded configuration combines faces at $\phi$ and $-\phi$. 
Analytical techniques developed for first-passage process with 
multiple adsorbing boundaries can be used to calculate the yield as a function of $N$.

\begin{acknowledgements}
We acknowledge financial support from the Portuguese
Foundation for Science and Technology (FCT) under
Contracts No. PTDC/FIS-MAC/28146/2017 (LISBOA--01--0145--FEDER--028146), UIDB/00618/2020, and
UIDP/00618/2020.
\end{acknowledgements}

\bibliographystyle{spphys}
\bibliography{Kirigami_Paper_ref}

\end{document}